\documentclass[aps,pre,twocolumn,groupedaddress]{revtex4-2}
\ifx\pdfoutput\undefined
\usepackage{graphicx}
\else
\usepackage{epstopdf}
\usepackage{epsfig}
\usepackage{float}
\usepackage{amssymb,amsmath,stmaryrd,tabularx}
\usepackage{wrapfig}
\usepackage{bm}
\usepackage{ulem}
\usepackage{xcolor}
\fi

\usepackage[center]{subfigure}

\begin{document}

\title{DNA Supercoiling Drives a Transition between Collective Modes of Gene Synthesis}

\author{Purba Chatterjee, Nigel Goldenfeld, and Sangjin Kim}

\affiliation{Department of Physics and Center for the Physics of Living Cells, University of Illinois at Urbana-Champaign, Loomis Laboratory of Physics, 1110 West Green Street, Urbana, Illinois 61801, USA\\
Carl R. Woese Institute for Genomic Biology, University of Illinois at Urbana-Champaign,
 1206 West Gregory Drive, Urbana, Illinois 61801, USA}

\date{\today}

\begin{abstract}

Recent experiments showed that multiple copies of the molecular machine RNA polymerase (RNAP) can efficiently synthesize mRNA collectively in the active state of the promoter. However, environmentally-induced promoter repression results in long-distance antagonistic interactions that drastically reduce the speed of RNAPs and cause a quick arrest of mRNA synthesis. The mechanism underlying this transition between cooperative and antagonistic dynamics remains poorly understood. In this Letter, we introduce a continuum deterministic model for the translocation of RNAPs, where the speed of an RNAP is coupled to the local DNA supercoiling as well as the density of RNAPs on the gene. We assume that torsional stress experienced by individual RNAPs is exacerbated by high RNAP density on the gene and that transcription factors act as physical barriers to the diffusion of DNA supercoils.
We show that this minimal model exhibits two transcription modes mediated by the torsional stress: a fluid mode when the promoter is active and a torsionally stressed mode when the promoter is repressed, in quantitative agreement with experimentally observed dynamics of co-transcribing RNAPs. Our work provides an important step towards understanding the collective dynamics of molecular machines involved in gene expression.
\end{abstract}

\maketitle

Transcription of the genome by the molecular machine RNA polymerase (RNAP) is the first step of gene expression and one of the key cellular processes. Transcription is considered to proceed through a collective mechanism: multiple RNAPs concurrently transcribing a gene increase their efficiency through cooperative interactions \cite{Epshtein2003,Epshtein2003a,Saeki2009,Jin2010,Kulaeva2010,Tantale2016,Le2018}. Mechanisms based on close-ranged interactions, such as ``push" \cite{Epshtein2003} or ``push-pull" \cite{Tantale2016,Heberling2016}, have been proposed to explain the cooperation. Also, theoretical models have been formulated to explain the emergence of such collective dynamics \cite{Galburt2011,costa2013,Heberling2016,Lesne2018,Belitsky2019}. These models generally predict that the speed of an RNAP (\lq\lq transcription elongation rate") increases with the density of RNAPs, and consequently, with the rate at which the promoter initiates transcription (RNAP loading). However, whether this scaling of the elongation rates with initiation rates holds for a wide range of promoter strengths and activities remains unclear.

A recent study on the transcription of \textit{lacZ} gene in the bacterium \textit{Escherichia coli} showed that in the active state of the promoter, the elongation rate remains constant for a large range of intermediate to high initiation rates, for which more than one RNAP transcribes the gene at the same time \cite{Kim2019}. More surprisingly, upon environmentally-induced promoter repression, the speed of RNAPs far downstream of the promoter is reduced drastically to values that are only a fraction of the speed of a single RNAP transcribing alone. This drastic decrease in RNAP speed is more pronounced if the repression happens while the first RNAP is about to finish transcription, which is contrary to intuition. The dynamics of a single RNAP is unaffected by promoter repression, suggesting that the observed drop in efficiency results from a collective antagonistic effect among multiple RNAPs.

Which mechanism of transcription regulation underlies this switch from cooperative to antagonistic dynamics? This switch is likely mediated by transcription-induced DNA supercoiling because it was observed in topologically constrained DNA templates, such as a plasmid and chromosome, but not in linear DNA, where two ends can freely rotate to dissipate DNA supercoils \cite{Kim2019}. Forward translation of an RNAP results in under-winding of the DNA behind (negative DNA supercoiling) and over-twisting of the DNA in front (positive DNA supercoiling) \cite{Liu1987}. Accumulation of these supercoils are known to slow down the RNAP due to torsional stress \cite{Rovinskiy2012,Ma2013,Chong2014}. However, positive and negative DNA supercoils can cancel commensurately between co-transcribing RNAPs \cite{Guptasarma1996,Koster2010,Liu1987} regardless of the distance between RNAPs (or RNAP density on the gene). Thus, elongation rates remain high independent of initiation rates as long as RNAP loading is uninterrupted. This cooperative dynamics can become antagonistic when loading is stopped (promoter repression): the last loaded RNAP slows down due to the accumulation of DNA supercoils behind it. With increasingly insufficient cancellation of supercoils downstream, other RNAPs on the gene progressively slow down, too. Although qualitatively able to capture the observed dynamics, this mechanism does not explain how RNAPs close to the end of a gene slow down so rapidly to only a quarter of a single RNAP speed upon promoter repression, without making as many DNA supercoils. The model also fails to explain why repressing the promoter closer to transcription completion of the first few RNAPs yields a more pronounced reduction in transcription efficiency.

The purpose of this paper is to introduce a minimal deterministic model for the expression of a typical gene regulated by transcription-induced DNA supercoiling. Our model is based on two novel hypotheses regarding the mechanism of torsional-stress generation during transcription. The first hypothesis is that the stress due to DNA supercoiling is exacerbated when there are many RNAPs on the gene. The second hypothesis posits that transcription factors (TFs), which bind near the promoter and block transcription initiation, act as physical barriers to the diffusion of DNA supercoils. They thus impose different conditions of torsional stress on the transcription dynamics of the active and repressed states of the promoter. These two hypotheses are crucial for our results. We show that this model results in two modes of transcription mediated by the torsional stress due to DNA supercoiling: a fluid mode when the promoter is active and a torsionally stressed mode when the promoter is repressed. Despite its simplicity, this minimal model accurately recapitulates the experimental observations of \cite{Kim2019} and is a step towards a semi-quantitative understanding of collective effects during gene expression.

\section{The Model}
\begin{figure}
\begin{center}
\includegraphics[scale=0.145]{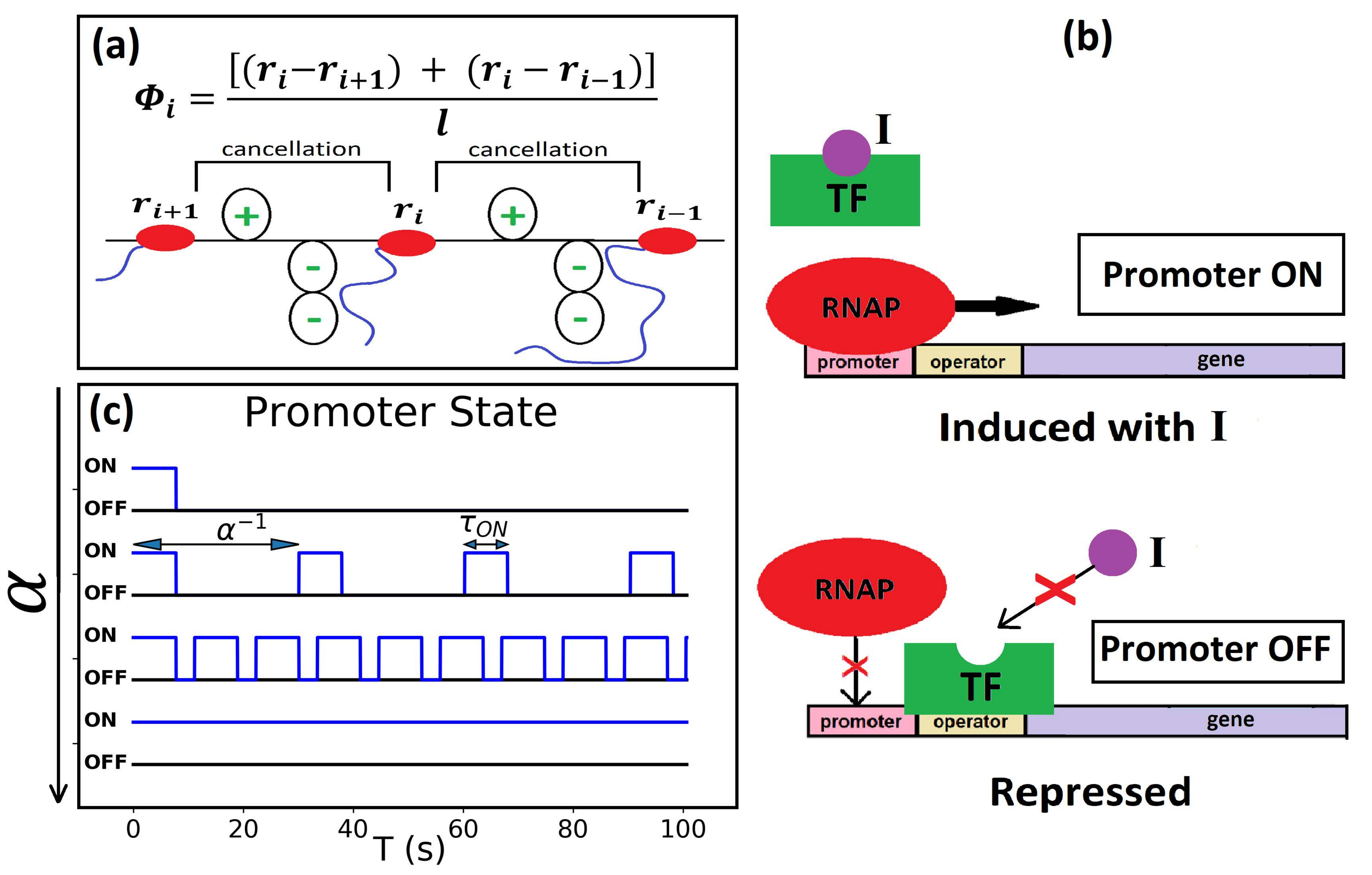}
\end{center}
\caption {\small{(color online) The model. (a) Local DNA supercoiling of the $RNAP_i$. (b) ON and OFF states of the promoter regulated by TF binding. (c) Time series of the promoter state for four different initiation rates $\alpha$.}}
    \label{fig1}
\end{figure}
We model the $i^{th}$ RNAP ($RNAP_i$) to be a point particle translocating on a gene of length $L$ with its position on the DNA given by $r_i$ and speed by $v_i=\dot{r}_i$. During transcription elongation, the bulky RNAP does not rotate with the DNA but twists the DNA to generate positive supercoils (PS) in front and negative supercoils (NS) behind \cite{Liu1987}. The DNA supercoiling $\phi_i$ experienced by $RNAP_i$ is formulated as the sum of the net NS behind it and the net PS in front of it after cancellations of DNA supercoils from adjacent RNAPs. We define
\begin{equation}\label{eq1}
\phi_i=\big[(r_i-r_{i+1})+(r_i-r_{i-1})\big]/l,
\end{equation}
where $l=10.5$ is the distance traversed by an RNAP before it makes one complete DNA supercoil (Fig.~\ref{fig1}(a)). 

The protocol of induction and repression for a typical gene is schematically illustrated in Fig.~\ref{fig1}(b). Before induction, TF is bound to the operator region and sterically hinders RNAPs from loading onto the promoter. The promoter can be turned ON by an inducer ($I$), which binds to TF and causes it to dissociate from the operator. The initiation rate $\alpha$ increases with the concentration of $I$. If $I$ is removed, the promoter can be turned OFF completely by repression. When repressed, TF remains bound and prevents any further loading. 

The promoter state for different initiation rates $\alpha$ is illustrated in Fig.~\ref{fig1}(c). We posit that three things happen in sequence whenever the promoter turns ON. First, TF is made to dissociate from the operator by I. This allows for the second event, the diffusion of NS behind the last loaded RNAP, removing torsional stress on this RNAP. Here, we are making the hypothesis that TF, which is also a bulky molecule, can physically block the diffusion of DNA supercoils \cite{Leng2002,Fulcrand2016}. We provide further support for this hypothesis in the discussion. Also, we assume that the diffusion takes place before the next RNAP loads based on the observation that the diffusion of DNA supercoils is very fast, about a $100$ times faster than RNAP dynamics \cite{Loenhout2012}. The third event is the loading of a new RNAP. The promoter remains ON for a duration $\tau_{ON}$, which is the average time taken by TF to rebind. As long as the promoter remains ON, no DNA supercoils accumulate behind the last loaded RNAP. When the promoter turns OFF, TF rebinds and blocks further dissipation of DNA supercoils and the loading of RNAPs till the next time the promoter turns ON. Repression at time $T_{stop}$ turns the promoter OFF completely thereafter and prevents any further RNAP loading. Note that we allow PS in front of $RNAP_1$ to diffuse downstream unhindered; however, the relaxation of this assumption does not change the main results of the model.

We further hypothesize that the torsional stress experienced by an RNAP depends on the amount of DNA supercoiling $\phi$ (local effect) as well as the number of RNAPs on the gene, $n$ (global effect). The latter is motivated by the idea that having many bulky RNAP molecules on the gene would make it harder to twist the DNA. The torsional stress experienced by $RNAP_i$, denoted by $\sigma_i$, is defined as
\begin{equation}\label{eq2}
\sigma_i(\phi_i,n)=(\gamma\phi_if(n))^b.
\end{equation}
Here, $f(n)$ is a monotonically increasing function of RNAP density $n$, and $\gamma=l/L$ is a normalization factor. The torsional stress $\sigma_i$ is taken to be a high odd power $b$ of $\phi_i$ to ensure that the stress is more pronounced at higher levels of DNA supercoiling. The speed of $RNAP_i$ decreases with increasing torsional stress as
\begin{equation}\label{eq3}
v_i(\sigma_i)=\frac{2v_0}{1+\exp(\beta\sigma_i)},
\end{equation}
where $v_0$ is the typical RNAP speed and $\beta$ is the rate of speed decay. A negative $\sigma_i$ indicates the presence of NS in \textit{front} of $RNAP_i$, which assists its speed \cite{Ma2013}. Thus, while the speed drops to zero for high positive stress, it doubles compared to $v_0$ for high negative stress. 

\section {Model Parameters and Methods}

\begin{figure}
\begin{center}
\includegraphics[scale=0.14]{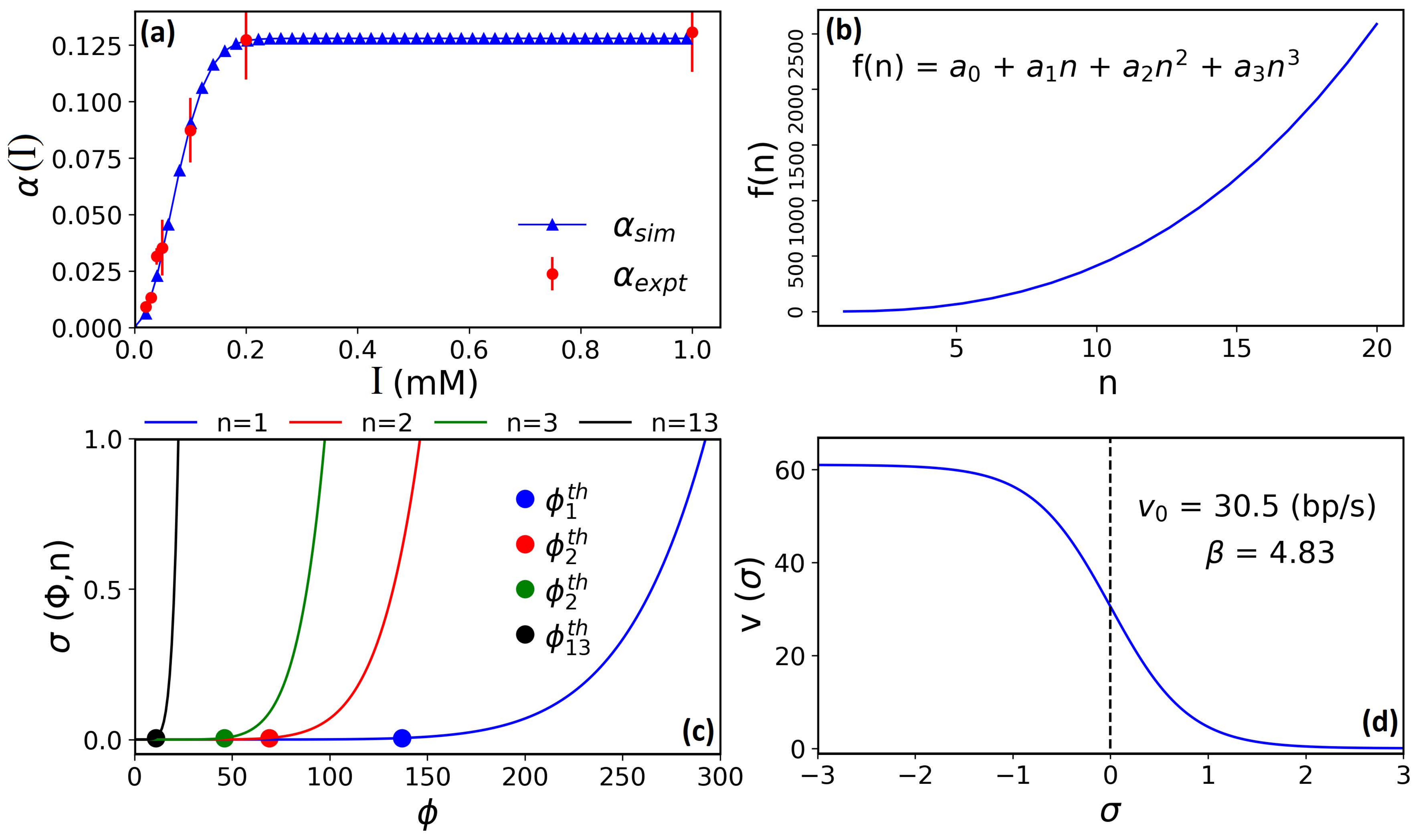}
\end{center}
\caption {\small{(color online) Model parameters. (a) Initiation rates $\alpha$ used in the simulations (blue triangles) based on experimentally observed initiation rates (red solid circles) in \cite{Kim2019}. (b) $f(n)$ as a function of $n$.\textbf{ $a_0 = 4.18$, $a_1= -6.16$, $a_2=2.74$, $a_3=0.25$}. (c) Torsional stress $\sigma$ as a function of DNA supercoiling $\phi$ for four different values of $n$. (d) Speed of an RNAP $v$ as a function of torsional stress $\sigma$.}}
    \label{fig2}
\end{figure}

As proof of concept, we apply our general model to the \textit{lac} operon in \textit{E. coli}, a paradigm of bacterial gene regulation, for which experimental results are available from \cite{Kim2019}. We focus on the transcription of \textit{lacZ}, the first gene in the \textit{lac} operon, with length $L=3072$ bp. The effective initiation rates $\alpha_{sim}$ used in the simulations were obtained from a fit to the observed $\alpha_{expt}$ as a function of the concentration of inducer (I), i.e., isopropyl $\beta$-D-$1$-thiogalactopyranoside (IPTG) used in \cite{Kim2019} (Fig.~\ref{fig2}(a)). The TF in question is the repressor LacI. $\tau_{ON}$ is assumed to be the same for all inducer concentrations and is taken to be the inverse of $\alpha_{max}$, the highest frequency of RNAP loading observed experimentally. We choose $f(n)$ to be a cubic polynomial for simplicity (Fig.~\ref{fig2}(b)) and set $b=7$. With these choices, even though the torsional stress $\sigma$ rises more steeply for larger $n$, there is always a threshold amount of supercoiling $\phi^{th}_n$ for each $n$, below which the torsional stress is extremely low (solid circles in Fig.~\ref{fig2}(c)). Fig.~\ref{fig2}(d) shows the dependence of RNAP speed $v$ on the torsional stress $\sigma$, where we have chosen $v_0=30.5$ bp/s following \cite{Kim2019}.

The simulation is run for the time it takes the first few RNAPs to complete transcription. We follow the prescription in \cite{Kim2019} to calculate the average elongation rate of the first RNAP ($RNAP_1$). When the promoter is active, we calculate the average elongation rate as $v_{ON}=L/T_1$, where $T_1$ is the time taken by $RNAP_1$ to complete transcription. When the promoter is repressed at time $T_{stop}$, we calculate the average elongation rate of $RNAP_1$ for the remaining portion of the transcript after repression, as $v_{OFF}=(L-v_{ON}T_{stop})/(T_1-T_{stop})$. The definition of $v_{OFF}$ assumes that the RNAPs move at a constant speed $v_{ON}$ till the promoter is repressed at $T=T_{stop}$. While this is true for intermediate and high initiation rates, it underestimates the actual position of the single RNAP case with a low initiation rate because the single RNAP shows continuous slowing down from start to finish \cite{Tripathi2021} (see Fig.~\ref{figA3a}). Nevertheless, we adhere to this definition of $v_{OFF}$ for comparison with the experimental results of \cite{Kim2019}.

\section{Simulation Results} 

\begin{figure}
\begin{center}
\includegraphics[scale=0.165]{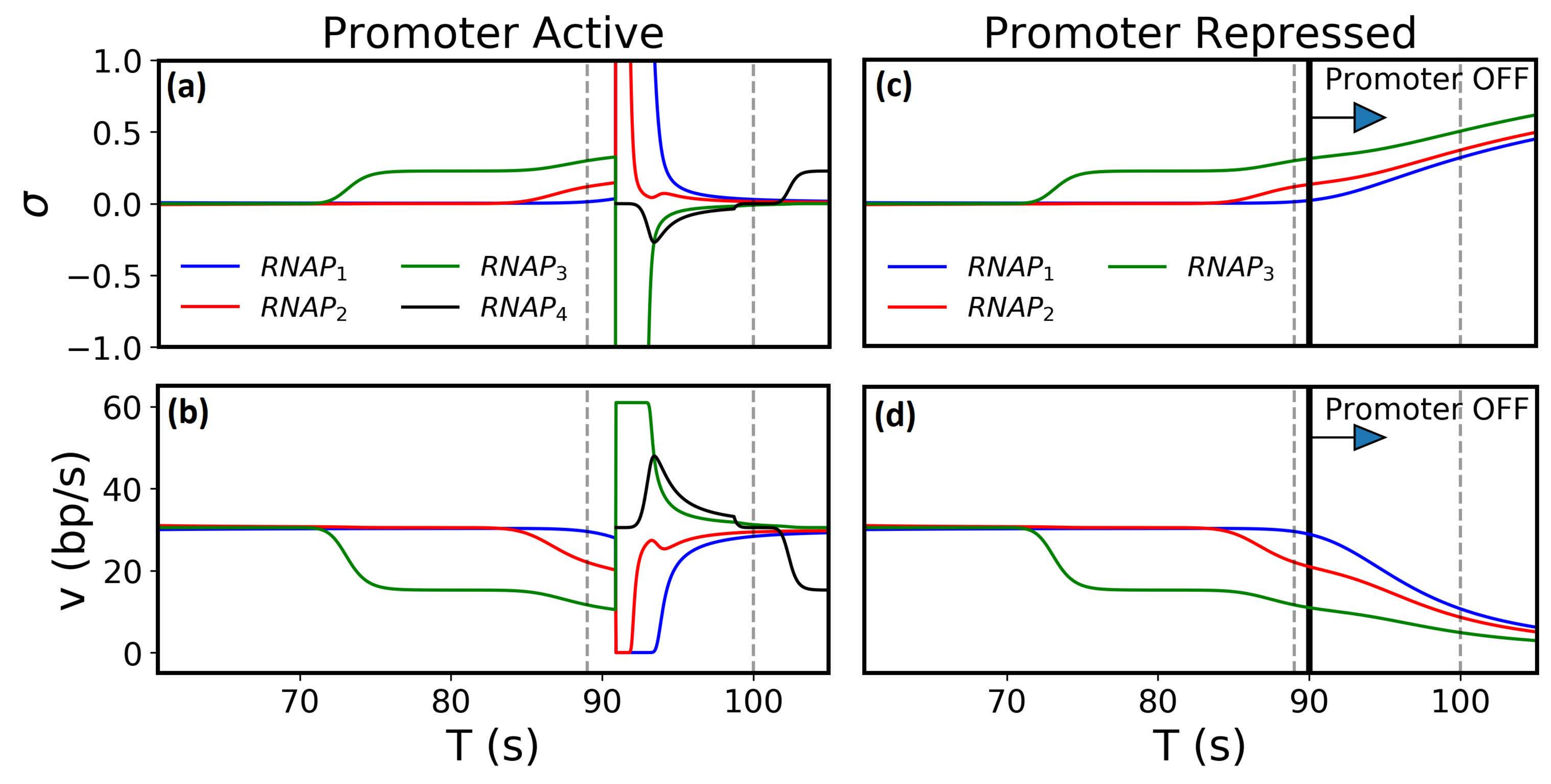}
\end{center}
\caption {\small{(color online) Time series of torsional stress $\sigma$ and speed $v$ of the first $3-4$ RNAPs for the case of active promoter (a-b) and promoter repression at $T=90$ s (c-d) at the intermediate initiation rate $\alpha=0.033$ s\textsuperscript{-1}. We only plot $\sigma \in(-1,1)$ for clarity, but the absolute value of torsional stress can increase beyond unity. Gray dashed lines demarcate the time range for Fig.~\ref{figA3b}.}}
    \label{fig3}
\end{figure}

Fig.~\ref{fig3}(a,b) shows the time series of torsional stress $\sigma$ and speed $v$ for the first $3-4$ RNAPs, at the intermediate initiation rate $\alpha=0.033$ s\textsuperscript{-1}. Note that in this regime of induction, the promoter cycles between ON and OFF states in the presence of $I$ (Fig.~\ref{fig1}(c) second from top). When the promoter is in the OFF state, DNA supercoils accumulate, but until $RNAP_3$ is loaded ($T=60$ s), the $n$-effect is not large enough to make the torsional stress significant (Fig.~\ref{fig2}(c)). TF rebinds after the $RNAP_3$ loads, and its blocking of DNA supercoil diffusion at the promoter leads to a sequential slowing down of RNAPs until the next TF dissociation event (at $T\approx91$ s). At this time, NS behind $RNAP_3$ dissipate, and all RNAPs on the gene can quickly equilibrate to the optimal speed $v_0$ by $T=100$ s, even in the presence of the newly loaded $RNAP_4$. This equilibration proceeds through the acceleration and deceleration of RNAPs reacting to the ambient torsional stress. It is detailed in Fig.~\ref{figA3b} for the time range demarcated by the dashed gray lines in Fig.~\ref{fig3}.

If the promoter is repressed at $T=90$ s (Fig.~\ref{fig3}(c,d)), TF remains bound ($RNAP_4$ does not load), and the speeds of $RNAP_1$, $RNAP_2$, and $RNAP_3$ continue to decrease. The speed of $RNAP_1$ is reduced to $10.7$ bp/s at $T=100$ s, and it decreases even further thereafter. Therefore, promoter repression in the model recapitulates the reduction in RNAP speeds observed in the experiments. (For the dynamics at low and high initiation rates, see Fig.~\ref{figA3a} and Fig.~\ref{figA3c}).

\begin{figure}
\begin{center}
\includegraphics[scale=0.16]{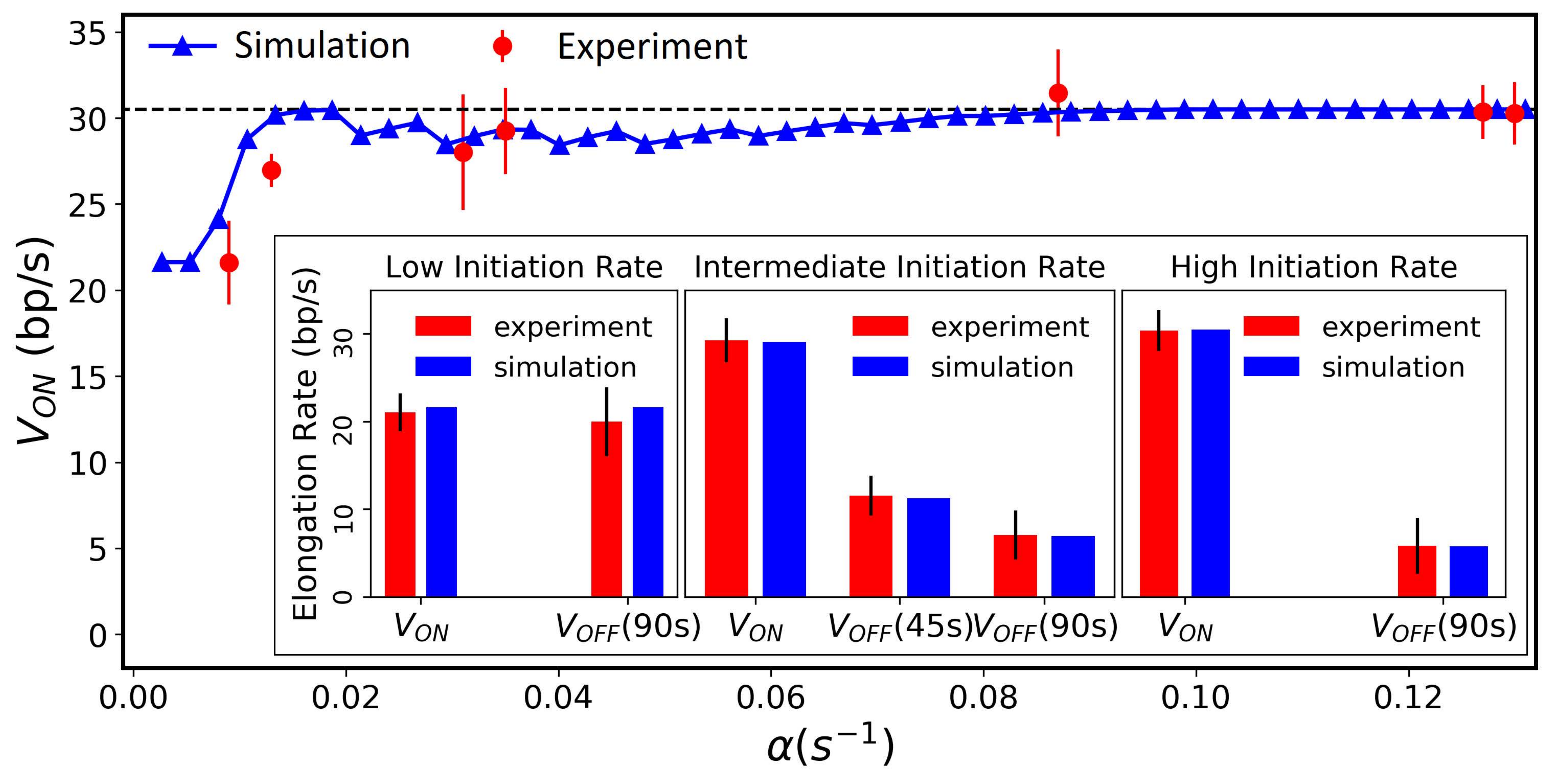}
\end{center}
\caption {\small{(color online) Comparison of our model (blue) with the experimental observations of \cite{Kim2019} (red) on the speed of RNAPs $v_{ON}$ as a function of initiation rates $\alpha$. The inset shows the effect of promoter repression. Low initiation rate represents $\alpha_{sim}=0.006$ s\textsuperscript{-1} ($\alpha_{expt}=0.009$ s\textsuperscript{-1}), yielding roughly a single RNAP on the gene at a given time. Intermediate and high initiation rates are from $\alpha_{sim}=0.033$ s\textsuperscript{-1} ($\alpha_{expt}=0.035$ s\textsuperscript{-1}) and $\alpha_{sim}=0.127$ s\textsuperscript{-1} ($\alpha_{expt}=0.127$ s\textsuperscript{-1}), respectively. At the intermediate initiation rate, repressing the promoter later at $T=90$ s produces lower average speeds than repression at $T=45$ s.}}
    \label{fig4_new}
\end{figure}

Fig.~\ref{fig4_new} shows the average elongation rate $v_{ON}$ vs initiation rate $\alpha$. For the low initiation rate ($\alpha_{sim}=0.006$ s\textsuperscript{-1}), there is only a single RNAP on the gene on average, and $v_{ON}=21.62$ bp/s, which is less than the typical speed $v_0=30.5$ bp/s. However, for a large range of higher $\alpha$, $v_{ON}\approx v_0$, independent of the initiation rate. The inset of Fig.~\ref{fig4_new} shows the average elongation rate $v_{OFF}$ upon promoter repression for three different initiation rates tested in the experiments. For the single RNAP case with a low initiation rate, promoter repression at $T=90$ s does not appreciably lower RNAP speed. For a high initiation rate, promoter repression at $T=90$ s causes speeds to drop to about a quarter of a single RNAP speed. Similarly, for an intermediate initiation rate, RNAP speeds show a sharp reduction after repression. Notably, the effect is greater if the promoter repression happens when the first RNAP is closer to the end of the gene (at $T=90$ s vs $T=45$ s), consistent with the experimental observations of \cite{Kim2019}.

\section{Discussion}

We summarize the proposed mechanism of transcription regulation as follows. When the promoter is active, the signal is to make as many transcripts as possible. The cancellation of DNA supercoils aids cooperative interactions between RNAPs, which move at high speed independent of initiation rates. When the promoter is repressed, the signal is to stop making transcripts. For higher RNAP densities, a stronger effect of this repression is needed in order to impose a quick arrest of further mRNA synthesis. Blocking the promoter results in the accumulation of NS behind RNAPs, followed by a drastic reduction of their speed. The antagonistic effect is greater for a higher density of RNAPs on the gene.

Our hypothesis that TFs regulate the diffusion of DNA supercoils is supported by experimental evidence of LacI functioning as a topological barrier to constrain DNA supercoils \cite{Leng2002,Fulcrand2016}. This novel hypothesis has important implications for modeling transcription dynamics in the genomic context, where genes may affect each other's transcription from a long distance via DNA supercoiling \cite{Meyer2014,Sobetzko2016}. For example, the diffusion of DNA supercoils or its lack thereof likely has important consequences for divergently transcribed genes commonly found in the genome \cite{Wei2011}. \cite{Kim2019} showed that the divergent expression of another gene, positioned upstream of \textit{lacZ}, reduces the transcription elongation rate of \textit{lacZ} in the case of the high initiation rate $\alpha=0.127$ s\textsuperscript{-1}. Moreover, this antagonistic effect was observed even when the two promoters are separated by as much as $2,400$ bp. This is entirely consistent with our model's prediction. For high initiation rates, the promoter is almost always ON (Fig.~\ref{fig1}(c) highest $\alpha$). As a result, NS generated by the transcription of a neighboring gene can diffuse in and reduce the speed of RNAPs transcribing \textit{lacZ}. 

To our knowledge, neither the role of TFs as a barrier to DNA supercoil diffusion, nor the dependence of torsional stress on the number of RNAPs on the gene, has been explored in existing theories of transcription based on DNA supercoiling \cite{Heberling2016,Brackley2016,Sevier2017}. However, we find that these assumptions play a very important role in producing the extremely contrasting RNAP dynamics between the active and repressed states of the promoter. As shown in Fig.~\ref{figA1a}-\ref{figA2}, relaxing either of these assumptions fails to produce the drastic slowdown of RNAPs after promoter repression, as experimentally observed in \cite{Kim2019}.
\begin{figure}
\begin{center}
\includegraphics[scale=0.15]{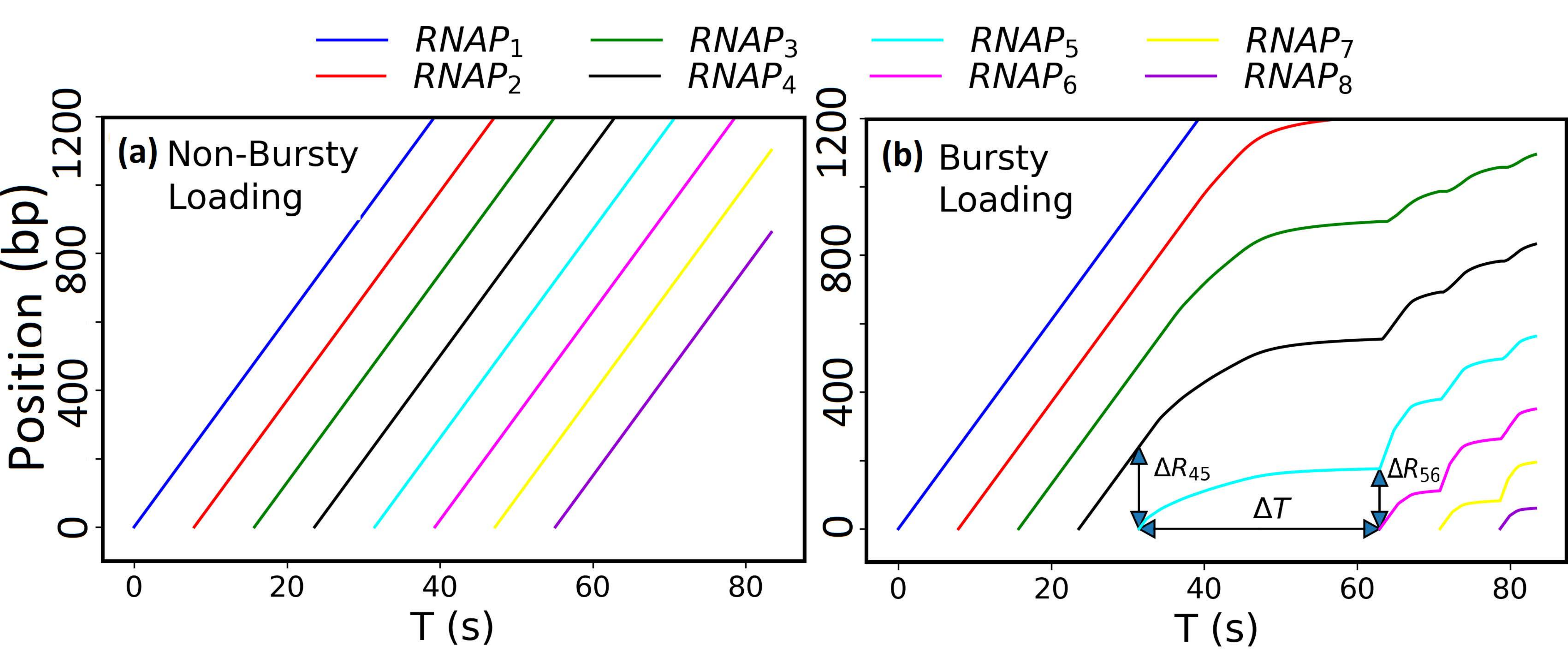}
\end{center}
\caption {\small{(color online) Translocation dynamics of RNAPs from nonbursty and bursty initiation. (a) Trajectories of RNAPs from nonbursty initiation at $\alpha=0.127$ s\textsuperscript{-1}. (b) Trajectories of RNAPs from bursty initiation, where $5$ RNAPs are loaded in a burst and $\Delta T >\alpha^{-1}$ is the duration between bursts.}}
    \label{fig5}
\end{figure}

In our model, RNAP loading is assumed to be at regular intervals, motivated by the absence of convoy formation (bursty transcription) under the experimental conditions of \cite{Kim2019}. However, \cite{Tantale2016} proposes that RNAPs loaded close to each other translocate at the same speed and travel as a convoy during elongation. To test this scenario, we explicitly modeled bursty initiation, with $5$ RNAPs loading in quick succession within a single burst and a longer duration between bursts. The results are shown in Fig.~\ref{fig5}(b) in comparison to nonbursty loading in Fig.~\ref{fig5}(a). We find that even though the loading of the last RNAP in a burst ($RNAP_5$) and the first RNAP in the next burst ($RNAP_6$) is separated by a longer time duration ($\Delta T$) than the loading of RNAPs within a burst(e.g., $RNAP_4$ and $RNAP_5$), the physical separation between them ($\Delta R_{56}$) is comparable to the separation between RNAPs in a particular burst ($\Delta R_{45}$). Here, the accumulation of NS behind the last RNAP in a burst leads to reducing its speed and thus the physical distance to the RNAP in the next burst. This prediction of our model suggests that convoy formation is hindered even with imposed bursty initiation. Additional mechanisms removing DNA supercoils may help maintain convoys after bursty transcription initiation.

In conclusion, we presented a continuum deterministic model for RNAP translocation affected by DNA supercoiling as well as RNAP density. A fluidic mode of transcription elongation was observed in a wide range of initiation rates because torsional stress remains low through supercoil cancellations. Our model showed that promoter repression can result in the torsionally stressed mode of elongation - and not merely due to the lack of supercoil cancellation. The accumulation of DNA supercoiling is exaggerated with bound TF and high RNAP density. Most importantly, the switch from cooperative to antagonistic RNAP dynamics upon promoter repression is mediated by purely mechanistic effects of transcription-induced DNA supercoiling.

After the completion of this work and during the preparation of this paper, we received a preprint from Tripathi, et al. \cite{Tripathi2021} which overlaps with some of the results we report.

This work was supported by the National Science Foundation Center for Physics of Living Cells (Grant No. NSF PHY-1430124). P.C. acknowledges the Drickamer Research Fellowship, $2020$. S.K. acknowledges support from the Searle Scholars Program.


\section*{Appendix}

\setcounter{equation}{0}
\setcounter{section}{0}
\setcounter{figure}{0}
\renewcommand{\theequation}{A\arabic{equation}}
\renewcommand{\thefigure}{A\arabic{figure}}
\renewcommand{\thesection}{A\arabic{section}}
\renewcommand{\thesubsection}{\arabic{subsection}}

In this appendix, we provide the details of RNAP dynamics for different initiation rates and promoter activities, i.e., active and repressed promoter (Sec.~\ref{sec1}). Furthermore, we try to relax the central hypotheses of our model and compare the results with experimental data (Sec.~\ref{sec2} and Sec.~\ref{sec3}). 

\section{Comparison of Dynamics For Active and Repressed Promoters}\label{sec1}

Here, we compare the torsional stress and speeds of the first few RNAPs for different promoter strength (low, intermediate, and high initiation rates) and activities (active and repressed).
\subsection{Low Initiation Rate}\label{sec1.1}
\begin{figure}[h]
\begin{center}
\includegraphics[scale=0.14]{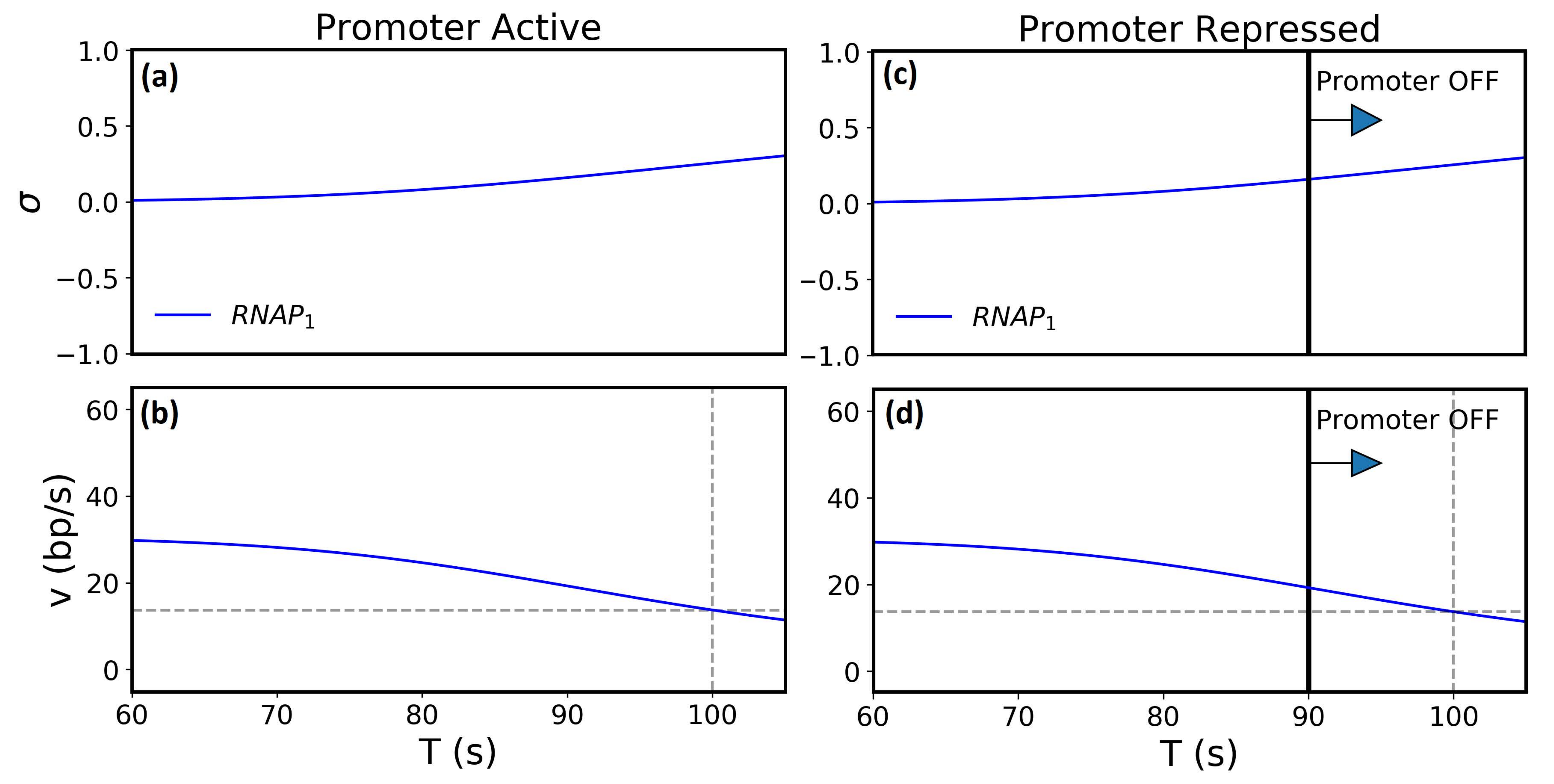}
\end{center}
\caption {\small{(color online) Time series of torsional stress $\sigma$ and speed $v$ of an RNAP at the low initiation rate $\alpha=0.006$ s\textsuperscript{-1} when the promoter is active (a-b) and when the promoter is repressed at $T=90$ s (c-d). The speed of the single RNAP continues to decrease even when the promoter is active (b), and the same RNAP dynamics is seen even when the promoter is repressed $T=90$ s (d).}}
    \label{figA3a}
\end{figure}
At the low initiation rate ($\alpha=0.033$ s\textsuperscript{-1}), there is only a single RNAP on the gene on average. Fig.~\ref{figA3a} shows the time series of torsional stress $\sigma$ and speed $v$ of the single RNAP for the active promoter (Fig.~\ref{figA3a}(a,b)) and for the promoter repressed at $T=90$ s (Fig.~\ref{figA3a}(c,d)). In both cases, the speed decreases continuously, such that at $T=100$ s we have $v=13.72$ bp/s for both active \textit{and} repressed promoters. The time taken by the single RNAP to complete transcription, $T_1$, is the same in both cases. Due to low $\alpha$, there is no upstream RNAP that can assist through DNA supercoil cancellation. Thus, even when the promoter is active, the dynamics of a single RNAP is subject to similar levels of torsional stress irrespective of the promoter state. 

Because the RNAP speed decreases continuously at low initiation rates, the definition of the average elongation rate after repression in \cite{Kim2019}, $v_{OFF}=(L-v_{ON}T_{stop})/(T_1-T_{stop})$, underestimates the position of the single RNAP upon repression (i.e., when $T = T_{stop}$). As a result, it predicts a higher $v_{OFF}$ than if we were to consider the actual position $r_1 (T_{stop})$. This is a limitation of the experiment, which does not track the position of the RNAPs. One has to assume a constant speed $v_{ON}$ till $T_{stop}$ in order to calculate the speed after repression, $v_{OFF}$.

Whenever the time taken to complete transcription with the active promoter ($T_{end}^{ON}$) is equal to that with the promoter repressed at $T_{stop}$ ($T_{end}^{OFF}$), this definition of $v_{OFF}$ always predicts $v_{OFF}=v_{ON}$, as seen by the experiments \cite{Kim2019} (Fig.~\ref{fig4_new}). That is, with $L=v_{ON}T_{end}^{ON}$, we have
\begin{align}\label{eq1.1}
v_{OFF}&=\frac{L-v_{ON}T_{stop}}{T_{end}^{OFF}-T_{stop}},\nonumber\\
&=\frac{v_{ON}(T_{end}^{ON}-T_{stop})}{T_{end}^{OFF}-T_{stop}},\nonumber\\
\implies v_{OFF}&=v_{ON},
\end{align}
for $T_{end}^{OFF}=T_{end}^{ON}$. The actual elongation rate after repression would be lower than that calculated in the experiments \cite{Kim2019} on average. However, we chose to adhere to this definition of $v_{OFF}$ for accurate comparison with the experimental results of \cite{Kim2019}. What is important to note is that the dynamics of the RNAP remains unaffected by the promoter state and that the time taken to complete transcription is the same for both active and repressed promoters (regardless of $T_{stop}$) when there is a single RNAP on the DNA. Thus, one should read the result $v_{ON}=v_{OFF}$ for the single RNAP case as $T_{end}^{OFF}=T_{end}^{ON}$, that is the time of transcription completion for a single RNAP is unaffected by active or repressed conditions of the promoter. 

\subsection{Intermediate Initiation Rate}\label{sec1.2}

\begin{figure}[h]
\begin{center}
\includegraphics[scale=0.14]{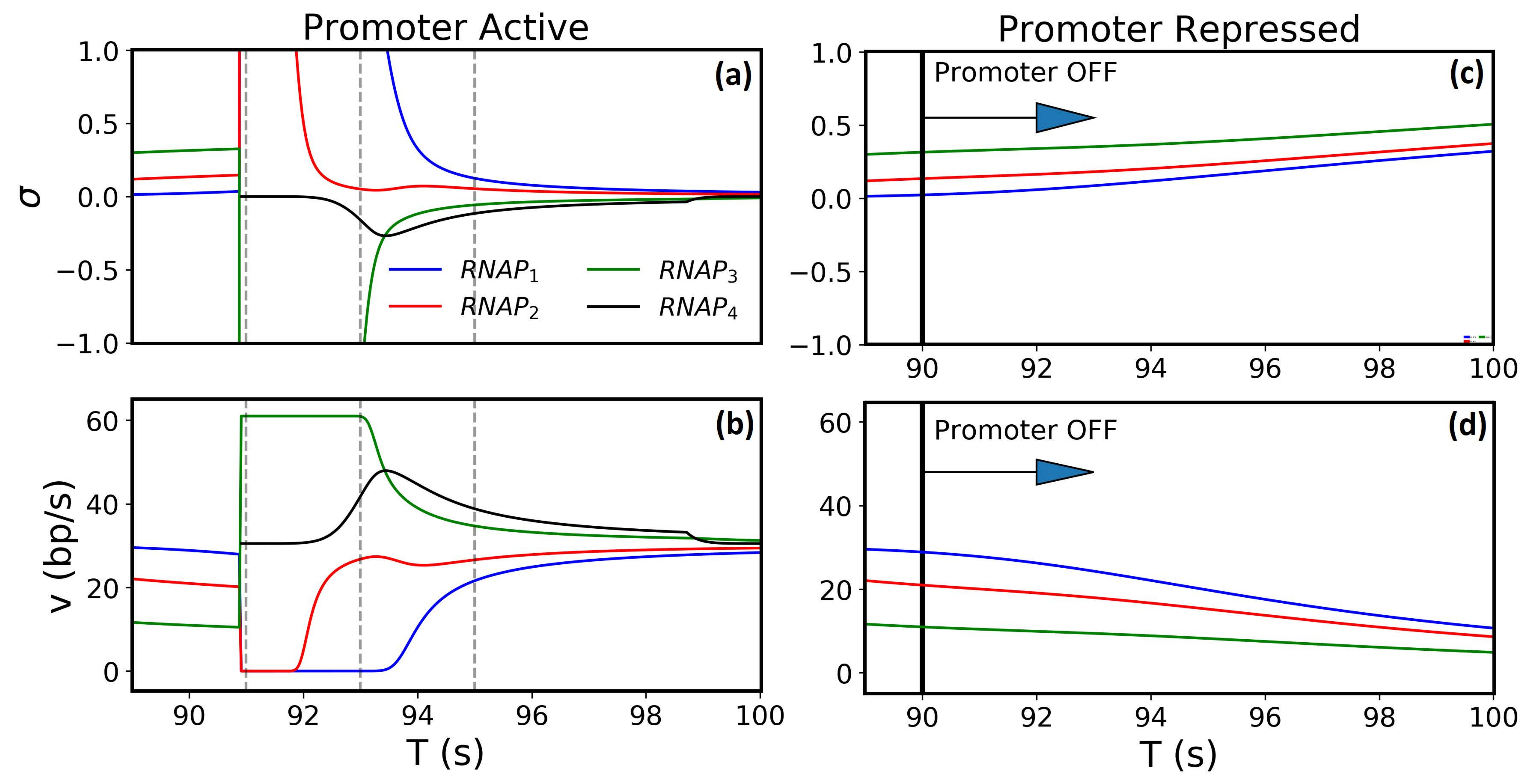}
\end{center}
\caption {\small{(color online) Time series of torsional stress $\sigma$ and speed $v$ of the first $3-4$ RNAPs from a promoter with the intermediate initiation rate $\alpha=0.033$ s\textsuperscript{-1}. (a-b) is when the promoter is continuously active, and (c-d) is when the promoter is repressed at $T=90$ s. This is a zoom-in version of Fig~\ref{fig3}. The dashed gray lines mark the time points discussed below. When the promoter is active, the first three RNAPs start at the typical speed $v_0$, but their speeds reduce due to NS accumulation. However, $RNAP_4$ loads at $T\approx 91$ s, and the resulting dynamics allows the speed of all four RNAPs to equilibrate to the typical speed $v_0$ by $T=100$ s. In contrast, with promoter repression at $T=90$ s, speeds of the three loaded RNAPs continue decreasing beyond $T=91$ s.}}
    \label{figA3b}
\end{figure}

Fig.~\ref{figA3b} shows the dynamics of RNAPs for the intermediate initiation rate $\alpha=0.033$ s\textsuperscript{-1}. We plotted the time series of torsional stress $\sigma$ and speed $v$ for the first $3-4$ RNAPs within the time range demarcated by gray dashed lines in Fig.~\ref{fig3}. At $T=89$ s, there are three RNAPs on the gene moving at speeds less than $v_0$, with $RNAP_3$ the slowest and $RNAP_1$ the fastest. $RNAP_3$ is slow because NS accumulates behind it while TF remains bound. There is a sequential slowing down of all downstream RNAPs starting from the promoter region due to insufficient cancellation of their NS by their slow upstream neighbor RNAPs. 

When the promoter stays active, TF dissociates for the next RNAP loading. In Fig.~\ref{figA3b}(a,b), TF dissociates at $T\approx 91$ s for the loading of $RNAP_4$, and there are a few consequences. In our model, when TF dissociates, the NS behind $RNAP_3$ diffuses out first prior to $RNAP_4$ loading, and hence, the remaining NS \textit{in front} of the $RNAP_3$ causes its speed to increase (green). As $RNAP_4$ loads, the speeds of $RNAP_1$ (blue) and $RNAP_2$ (red) fall, owing to the increased difficulty of translocating by overtwisting the DNA with an additional RNAP on the gene. The fast-moving $RNAP_3$ can cancel supercoils ahead more efficiently, so it speeds up $RNAP_2$ (see $T\approx 92$ s). $RNAP_4$ (black) initially speeds up right after loading because it has NS ahead, owing to the high speed of $RNAP_3$. Thus, at $T=93$ s, we see both $RNAP_2$ and $RNAP_4$ accelerating. Soon after, at $T=95$ s, $RNAP_3$ and $RNAP_4$ start slowing down due to NS accumulation behind them, and at the same time, $RNAP_1$ and $RNAP_2$ are speeding up due to better cancellation of their NS. Eventually, at around $T=100$ s, all four RNAPs have once again settled to the typical speed $v_0$. 

TF rebinds at some point after $RNAP_4$ loading and blocks supercoil diffusion. As more NS accumulate behind $RNAP_4$, it starts to slow down, re-initiating a sequential decrease in the speeds of downstream RNAPs. However, as long as loading is uninterrupted (active promoter), RNAPs can always equilibrate to the optimal speed. In contrast, when the promoter is repressed at $T=90$ s (Fig.~\ref{figA3b}(c,d)), $RNAP_4$ does not load, and the speeds of $RNAP_1$, $RNAP_2$, and $RNAP_3$ continue to decrease. At $T=100$ s, the speed of $RNAP_1$ reduces to $10.7$ bp/s, as compared to $28.4$ bp/s when the promoter remains active. The RNAP slows down even further after $T=100$ s to finally record an average elongation rate $v_{OFF} = 6.94$ bp/s, as shown in Fig.~\ref{fig4_new}. 

\subsection{High Initiation Rate}\label{sec1.3}

\begin{figure}[h]
\begin{center}
\includegraphics[scale=0.14]{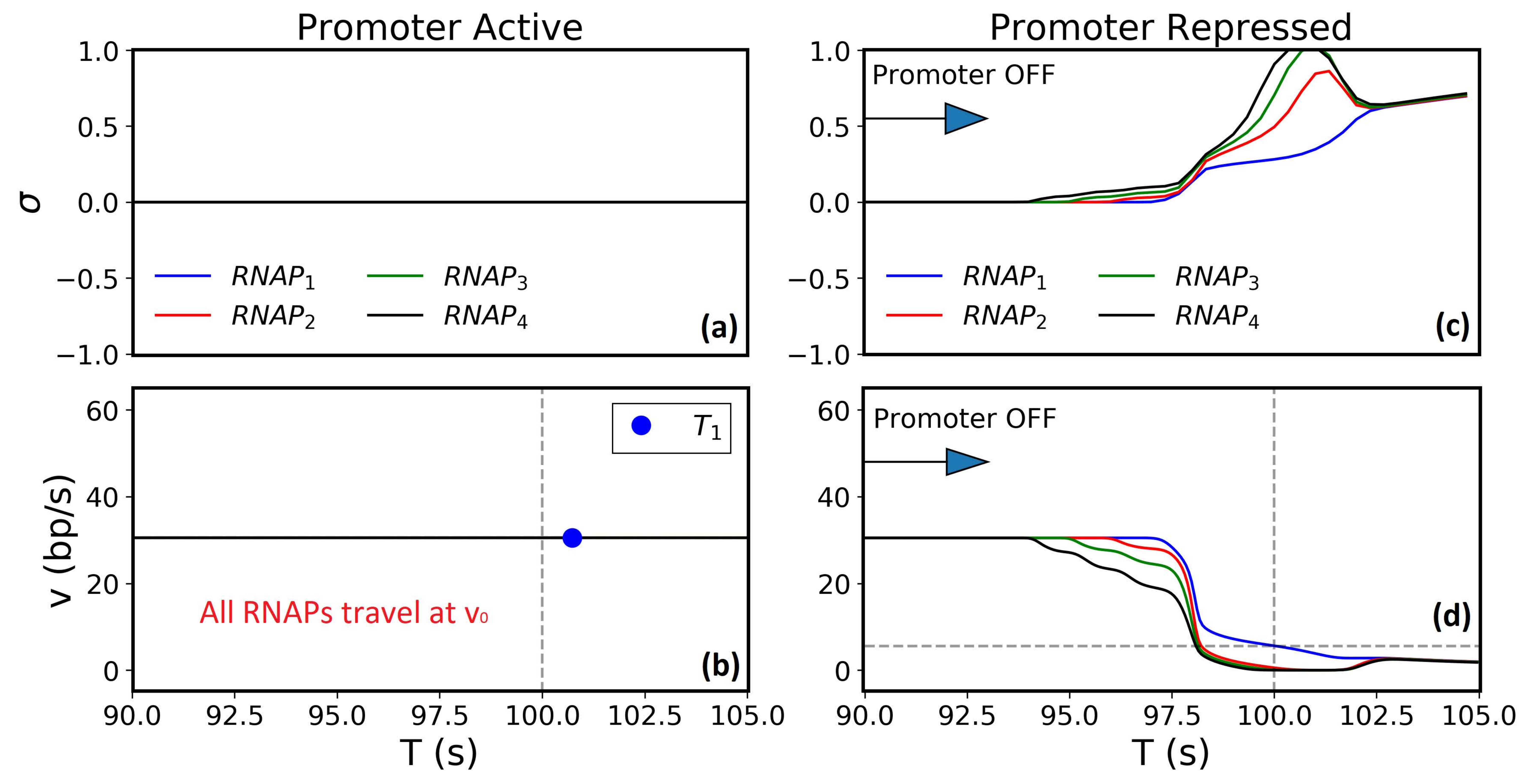}
\end{center}
\caption {\small{(color online) Time series of torsional stress $\sigma$ and speed $v$ of the first four RNAPs in the case of uninterrupted loading (a-b) and promoter repression at $T=90$ s (c-d) for the high initiation rate $\alpha=0.127$ s\textsuperscript{-1}. When the promoter remains active, all RNAPs travel at the typical speed $v_0$. In contrast, after promoter repression at $T=90$ s, speeds of the first four RNAPs suddenly reduce drastically over a short period of time as the torsional stress crosses threshold values.}}
    \label{figA3c}
\end{figure}

At a high initiation rate ($\alpha=0.127$ s\textsuperscript{-1}) of the active promoter, TF dissociates frequently, and the promoter is almost always ON (see Fig.~\ref{fig1}(c) for the highest $\alpha$). As such, NS do not accumulate behind the last loaded RNAP as long as the promoter is active. Fig.~\ref{figA3c} shows the time series of torsional stress $\sigma$ and speed $v$ of the first four RNAPs for the active promoter (Fig.~\ref{figA3c}(a,b)) and for the promoter repressed at $T=90$ s (Fig.~\ref{figA3c}(c,d)). It is clear that for the active promoter, there is negligible torsional stress throughout, and all RNAPs travel at the typical speed $v_0$. In Fig.~\ref{figA3c}(b), we have marked $T_1 =100.73$ s, the time taken for the first RNAP to complete transcription. In contrast, when the promoter is repressed at $T=90$ s, we see a drastic reduction of RNAP speeds over a very short period of time. For example, the speed of the first RNAP reduces to $v=5.57$ bp/s at $T=100$ s. This reduction is caused by TF binding at $T_{stop}$, which prevents both further initiation as well as the diffusion of NS produced by the last loaded RNAP. Moreover, because there are approximately $n=12$ RNAPs on the gene when the promoter is repressed, the accumulation of a very small amount of supercoiling is sufficient to increase the torsional stress beyond threshold values. Thus, promoter repression in the model recapitulates the drastic reduction in RNAP speeds observed in the experiments \cite{Kim2019}. A greater reduction is expected for larger RNAP densities on the gene, i.e for higher initiation rates, suggesting that this antagonistic effect is another group effect of RNAPs.

\section{Torsional Stress Independent of RNAP Density }\label{sec2}
In our model, we hypothesize that the presence of many RNAPs on the gene exacerbates the torsional stress by making the DNA more difficult to overtwist. The dependence of the torsional stress $\sigma$ on the RNAP density $n$ is encoded by the function $f(n)$, shown in Fig.~\ref{fig2}(b). To relax this assumption in our model, we consider the situation where $\sigma$ is independent of $n$, i.e. $f(n)=1$. In the following subsections, we examine $f(n)=1$ under three different scenarios related to supercoil diffusion at the promoter.

\subsection{TF Blocks NS Diffusion in its Bound State}\label{sec2.1}
\begin{figure}[h]
\begin{center}
\includegraphics[scale=0.14]{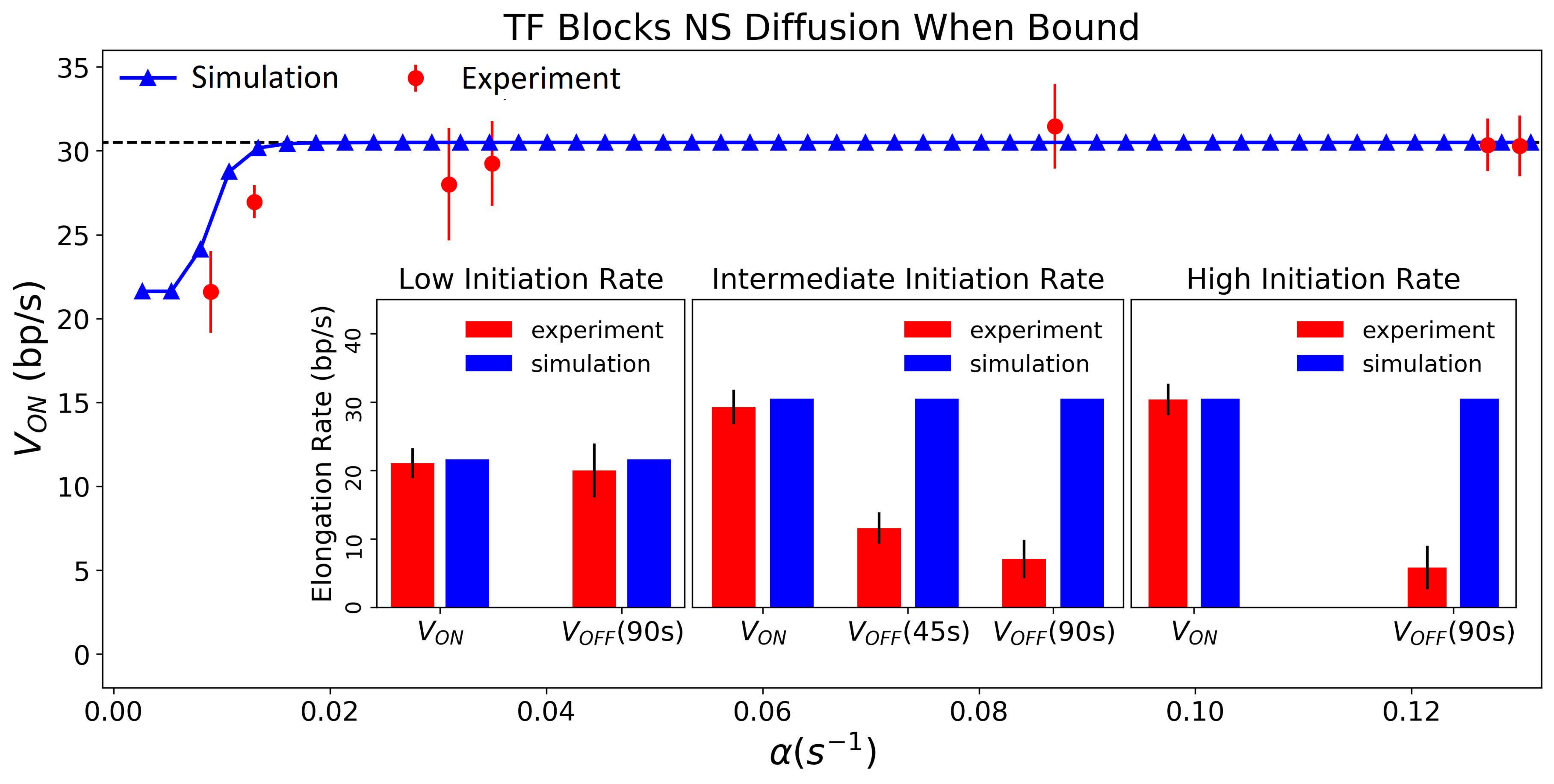}
\end{center}
\caption {\small{(color online) Scenario with $f(n)=1$ and TF blocking NS diffusion in its bound state. $v_{ON}$ is low for a single RNAP ($\alpha_{sim}=0.006$ s\textsuperscript{-1}), but it remains high independent of initiation rates for a large range of $\alpha$. The inset shows the effect of promoter repression. Promoter repression at $T=45$ s or $T=90$ s does not appreciable change $v_{OFF}$ from $v_{ON}$ for low ($\alpha_{sim}=0.006$ s\textsuperscript{-1}, $\alpha_{expt}=0.009$ s\textsuperscript{-1}), intermediate ($\alpha_{sim}=0.033$ s\textsuperscript{-1}, $\alpha_{expt}=0.035$ s\textsuperscript{-1}), and high ($\alpha_{sim}=0.127$ s\textsuperscript{-1}, $\alpha_{expt}=0.127$ s\textsuperscript{-1}) initiation rates.}}
    \label{figA1a}
\end{figure}

Fig.~\ref{figA1a} shows elongation rates of various conditions assuming that DNA-bound TF blocks NS diffusion (as in our main model) but $f(n)=1$. In the active state of the promoter, we once again see high elongation rates independent of initiation rates for intermediate to high $\alpha$, whereas the single RNAP case at low initiation rates (e.g., $\alpha=0.006$ s\textsuperscript{-1}) has a lower speed. However, for all initiation rates, promoter repression at neither $T=45$ s nor $T=90$ s shows any change in elongation rates $v_{OFF}$ from their $v_{ON}$ values (Fig.~\ref{figA1a} (inset)). This is in contrast to the experimental observation of \cite{Kim2019} that promoter repression causes a large reduction in RNAP speeds. Thus, without the dependence of torsional stress on RNAP density, even with TF blocking NS diffusion when bound, we cannot reproduce the observed switch from cooperative to antagonistic collective dynamics of RNAPs upon promoter repression.

\subsection{TF Never Blocks NS Diffusion}\label{sec2.2}
\begin{figure}[h]
\begin{center}
\includegraphics[scale=0.14]{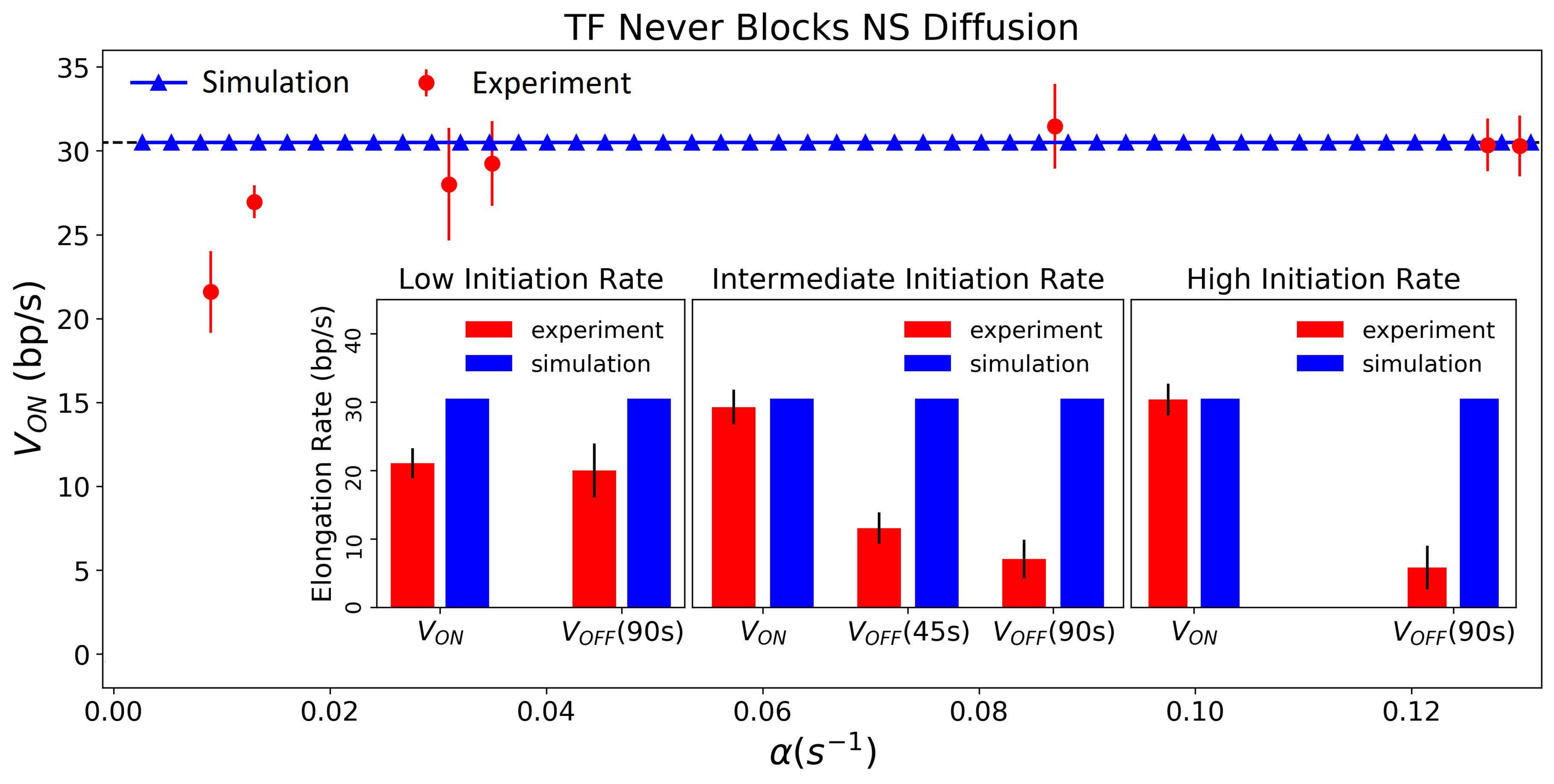}
\end{center}
\caption {\small{(color online) Scenario with $f(n)=1$ and TF never blocking NS diffusion. $v_{ON}$ is high and independent of initiation rates for all $\alpha$, even for a single RNAP ($\alpha_{sim}=0.006$ s\textsuperscript{-1}). The inset shows the effect of promoter repression. Promoter repression at $T=45$ s or $T=90$ s does not appreciable change $v_{OFF}$ from $v_{ON}$ for low ($\alpha_{sim}=0.006$ s\textsuperscript{-1}, $\alpha_{expt}=0.009$ s\textsuperscript{-1}), intermediate ($\alpha_{sim}=0.033$ s\textsuperscript{-1}, $\alpha_{expt}=0.035$ s\textsuperscript{-1}), and high ($\alpha_{sim}=0.127$ s\textsuperscript{-1}, $\alpha_{expt}=0.127$ s\textsuperscript{-1}) initiation rates.}}
    \label{figA1b}
\end{figure}

Fig.~\ref{figA1b} shows elongation rates for the case where $f(n)=1$ and TF never blocks NS diffusion. In other words, NS always diffuse out. This situation could arise in the case of a linearized plasmid that always allows supercoil dissipation through its free ends or in the case where TF is a comparatively smaller molecule and cannot constrain supercoils. In the active promoter, we see high elongation rates independent of initiation rates for \textit{all} $\alpha$. Even a single RNAP ($\alpha=0.006$ s\textsuperscript{-1}) transcribes at the optimal speed. This is contradictory to the experimental observation that co-transcribing RNAPs can collectively increase their elongation rates in comparison to a single RNAP. Additionally, like in Sec.~\ref{sec2.1}, promoter repression at $T=45$ s or at $T=90$ s does not show any change in elongation rates for any initiation rate (inset in Fig.~\ref{figA1b}). Thus, with the torsional stress independent of RNAP density ($f(n)=1$) and with TF unable to block NS diffusion even when bound, we cannot reproduce either the collective or the antagonistic dynamics of RNAPs observed in \cite{Kim2019}.

\subsection{No NS Diffusion}\label{sec2.3}
\begin{figure}[h]
\begin{center}
\includegraphics[scale=0.14]{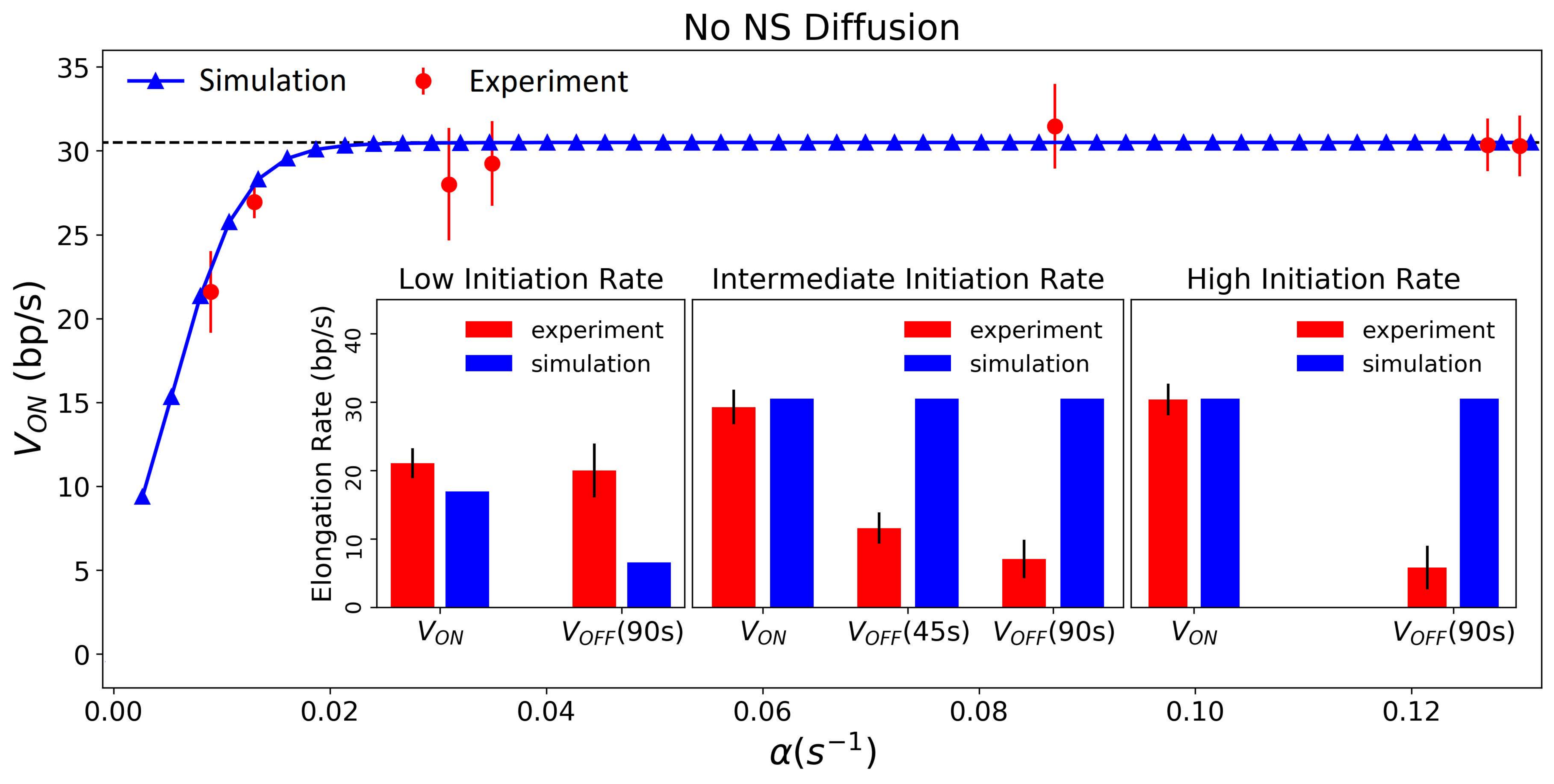}
\end{center}
\caption {\small{(color online) Scenario with $f(n)=1$ and no NS diffusion. The main panel shows that $v_{ON}$ is low for low initiation rates and high and independent of the initiation rates for intermediate and high $\alpha$. The inset shows the effect of promoter repression. Promoter repression at $T=45$ s or $T=90$ s does not appreciable change $v_{OFF}$ from $v_{ON}$ for intermediate ($\alpha_{sim}=0.033$ s\textsuperscript{-1}, $\alpha_{expt}=0.035$ s\textsuperscript{-1}) and high ($\alpha_{sim}=0.127$ s\textsuperscript{-1}, $\alpha_{expt}=0.127$ s\textsuperscript{-1}) initiation rates. However, the elongation rate of a single RNAP at the low initiation rate ($\alpha_{sim}=0.006$ s\textsuperscript{-1}, $\alpha_{expt}=0.009$ s\textsuperscript{-1}), reduces drastically from $v_{ON}$ for promoter repression at $T=90$ s.}}
    \label{figA1c}
\end{figure}

Fig.~\ref{figA1c} shows elongation rates for $f(n)=1$ and no NS diffusion. This case explores the scenario where TF binding or unbinding only affects RNAP loading but is irrelevant to the torsional stress. This can also be considered as the general case with a bulky molecule always bound to the DNA upstream of the promoter, which does not affect RNAP loading but blocks NS diffusion. 
Fig.~\ref{figA1c} shows that elongation rates are low for low initiation rates but high and independent of $\alpha$ for intermediate to high initiation rates of active promoters. This agrees with the observations of \cite{Kim2019}. However, promoter repression at $T=45$ s or at $T=90$ s does not show any change in the elongation rates for intermediate and high initiation rate (Fig.~\ref{figA1c} (inset)), contrary to \cite{Kim2019}. Another deviation from experimental data is that a single RNAP slows down after promoter repression at $T=90$ s. The single RNAP is slower in the absence of NS diffusion and requires another RNAP to load and relieve its torsional stress through cancellations before it can complete transcription. If promoter repression is at such a time that blocks this second RNAP from loading, it would result in a drastic decrease in the RNAP speed (Fig.~\ref{figA1c} for a low initiation rate $\alpha=0.006$ s\textsuperscript{-1}). Thus, without $n$ dependence and NS diffusion, we cannot capture the negative effect of promoter repression on the co-transcribing RNAPs or a single RNAP dynamics that are independent of the promoter state. 

\section{Supercoil Diffusion Not Blocked by TF }\label{sec3}
In our main model, we hypothesize that the presence and absence of TF on the DNA imposes different conditions of torsional stress on the transcription elongation dynamics. When TF is bound, it blocks NS diffusion and constrains them between itself and the last loaded RNAP. Unbinding of TF immediately results in the dissipation of this torsional stress, and we say that the NS behind the last loaded RNAP can diffuse out and not affect its speed. Keeping the $n$ dependence of the torsional stress identical to the main text (i.e., $f(n)$ as in Fig.~\ref{fig2}(b)), we tried to relax this assumption in two ways - first by looking at the scenario where NS always diffuse out (independent of TF binding as in Sec.~\ref{sec2.2}) and second by considering the case where NS never diffuse out (as in Sec.~\ref{sec2.3}). 

\begin{figure}[h]
\begin{center}
\includegraphics[scale=0.14]{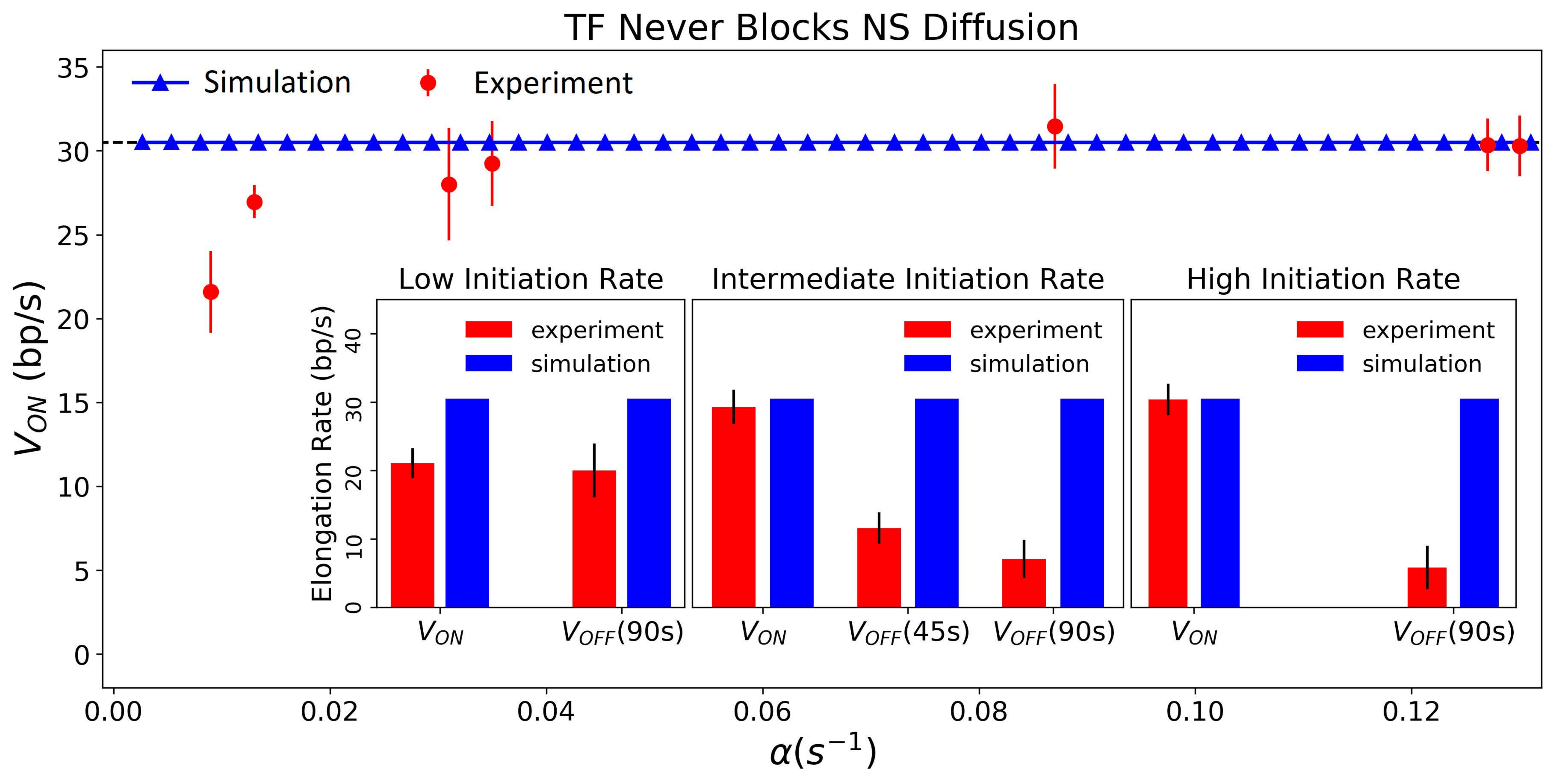}
\end{center}
\caption {\small{(color online) Scenario with $f(n)$ as shown in Fig.~\ref{fig2}(b) and TF never blocking NS diffusion. $v_{ON}$ is high and independent of initiation rates for all $\alpha$, even for a single RNAP ($\alpha_{sim}=0.006$ s\textsuperscript{-1}). The inset shows the effect of promoter repression. Promoter repression at $T=45$ s or $T=90$ s does not appreciable change $v_{OFF}$ from $v_{ON}$ for low ($\alpha_{sim}=0.006$ s\textsuperscript{-1}, $\alpha_{expt}=0.009$ s\textsuperscript{-1}), intermediate ($\alpha_{sim}=0.033$ s\textsuperscript{-1}, $\alpha_{expt}=0.035$ s\textsuperscript{-1}), and high ($\alpha_{sim}=0.127$ s\textsuperscript{-1}, $\alpha_{expt}=0.127$ s\textsuperscript{-1}) initiation rates.}}
    \label{figA2}
\end{figure}

Fig.~\ref{figA2} explores the first case, where TF never blocks NS diffusion but the torsional stress depends on RNAP density. The results are identical to the $f(n)=1$ case shown in Fig.~\ref{figA1b}: a fast elongation rate is maintained for \textit{all} initiation rates for both active \textit{and} repressed promoters. This result suggests that without TF blocking NS diffusion (as in a linearized plasmid or less massive TF), increasing torsional stress with RNAP density $f(n)$ is not sufficient to capture the observed cooperative dynamics of RNAPs for the active promoter nor the antagonistic dynamics upon promoter repression.

In the second case, where there is no NS diffusion but with the original $f(n)$ as in Fig.~\ref{fig2}(b), elongation rates drop to zero for all initiation rates and promoter states. This does not exclude the possibility that we can find another set of ($f(n), v_0, \beta$) that results in qualitative agreement with experimental observations, even with no NS diffusion. However, our efforts to find such a set of parameters revealed that under no conditions can we simultaneously capture two different phenomena: (i) high elongation rates independent of initiation rates (and hence of RNAP density) in the active state of the promoter and (ii) drastic slow-down upon promoter repression, with lower $v_{OFF}$ for a larger number of RNAPs on the gene. This result implies that LacI, the TF used in the experimental study \cite{Kim2019}, controls not only RNAP loading events but also the torsional stress of the elongation complexes depending on the ON and OFF states of the promoter. It remains to be tested whether this new role of TF can be found in other TFs or DNA-binding proteins, such as histones in eukaryotic cells.

\bibliographystyle{apsrev4-2}
\bibliography{RNAP_SC}
\end{document}